\documentclass[conference]{IEEEtran}

\usepackage[table]{xcolor}
\usepackage{calc} % for height calculateing
\usepackage{multirow} % Not sure what this does...
\usepackage{adjustbox}
\usepackage{cite}
\usepackage[caption=false]{subfig}

\definecolor{MNC}{HTML}{F0FF42}
\definecolor{WNC}{HTML}{3A00E0}
\definecolor{NNC}{HTML}{3DEEFF}
\definecolor{MTR}{HTML}{000000}
\definecolor{WTR}{HTML}{F50700}
\definecolor{NTR}{HTML}{FF8080}
\definecolor{MVS}{HTML}{70FFD0}
\definecolor{WVS}{HTML}{F53DFF}
\definecolor{NVS}{HTML}{FFBDFC}

\begin{document}

\author{

\IEEEauthorblockN{John Clements, Babak Farzad and Henryk Fuk\'s}
\IEEEauthorblockA{
Department of Mathematics and Statistics\\
Brock University\\
St. Catharines, Ontario, Canada\\
Email: jsclements642@gmail.com, bfarzad@brocku.ca, hfuks@brocku.ca
}
}

\title{ \huge \bfseries Dynamics of large scale networks following a merger}
%\thanks{ Faculty of Mathematics and Science, Brock University\\ St. Catharines, Ontario}

\maketitle

\begin{abstract}
We study the dynamic network of relationships among avatars in the massively multiplayer online game Planetside 2.
In the spring of 2014, two separate servers of this game were merged, and as a result, two previously distinct networks were combined into one. We observed the evolution of this network in the seven month period following the merger and report our observations.
We found that some structures of original networks persist  in the combined network for a long time after the merger. As the original avatars are gradually removed, these structures slowly dissolve, but they remain observable for a surprisingly long time. 
We present a number of visualizations illustrating the post-merger dynamics and discuss time evolution of selected quantities
characterizing the topology of the network.
\end{abstract}


\section{Introduction}
Among various types of large-scale networks which have been extensively studied in recent years \cite{lazer_life_2009}, online social networks 
received a lot of attention \cite{kwak_fragile_2011,szell_measuring_2010,son_analysis_2012}. Although numerous features of their dynamics have been investigated in great details,
not much is known about phenomena which one could call ``rare events''. One of such rare events is a merger of two networks, which will be the subject of this article.

Thanks to several months of advanced warning about the coming merger of two servers in the massively multiplayer online game (MMOG) PlanetSide 2, we were able to observe such an event, capture the relevant data, and perform some analysis. To our knowledge, 
this is the first study of a server merger reported in the literature.

Before we continue, we need to explain what a server merger means in the context of an MMOG. A server is a self-contained instance of the game world with its own and unique set of avatars. The interaction between servers varies greatly depending on the game but is very limited in Planetside~2, where there are no server transfers and no interactions beyond private messages and friendship links.
These MMOGs are designed for a large number of simultaneous players, and if the number of players drops below a certain point,
the administrators merge servers. There is no good real world analog for this sort of event, because in real life the mass transfer of a large population to another place would be accompanied by major disruptions of the associated social network. 
However, in a MMOG this process is almost completely ``painless'', as the two original social structures, fully intact, are simply placed on the same server. The closest real world analog would  probably be a merger of two social clubs or perhaps a business merger.

\section{Planetside 2 Data}
Our data consists of weekly snapshots of the network of friendships between the avatars in the computer game Planetside 2. Each of these snapshots is recorded as a graph with avatar being a node and unweighted undirected edges representing the  links between friends. Additional information such as avatar names, faction, time played, and other statistical data were collected on each avatar. 
Before the merger, we collected data from two servers to be combined, named Mattherson and Waterson. After the merger, we
collected data from the new combined server named Emerald, and, for control purposes, from two other servers named Connery and Miller.

The data were collected from the Sony Online Entertainment (SOE, http://census.soe.com) {census API} using our crawling algorithm. The API lets users query the SOE databases on several of the company's games, and has been used in studies of earlier games such as Everquest II \cite{shen_virtual_2012} as well as in the study of the outfit\footnote{In game communities, \emph{outfits} are formed by players, for organization and socialization. This is the same as clans or guilds in other MMOGs.} structure of Planetside 2 \cite{poor_collaboration_2014}. Since we were asked not to query out of date avatars, our crawler only follows links to avatars online in the last 44 days and disregards the rare links leading to different servers.

The data obtained from game servers were stored in SQLite databases. The data for a snapshot is recorded as a table of edges and a table of vertices and their attributes. The following items are included in the avatar attribute table of each snapshot.
\begin{itemize}
\item \emph{Id:} the unique identification of an avatar.
\item \emph{Name:}  the avatar's display name. Unlike the Id, the name can be changed (for a fee).
\item \emph{Battle rank (Br):} the avatar's ``battle rank'' is the Planetside 2 term for a level.  Battle ranks are capped at 100. 
\item \emph{Faction:}  each avatar chooses to belong to one of the three factions,  the New Conglomerate (NC), Terrain Republic (TR) and Vanu Sovereignty (VS). This choice cannot be changed.
\item \emph{Outfit:} a 2 to 4 letter name displayed in front of the avatar's name for every outfit member. While each tag is unique, they may be changed.  
\end{itemize}
It is important to keep in mind the distinction between players and avatars. The Census API only provides information on avatars, while the player information remains private. This means that one player could have many avatars, and the API can not directly identify which avatars correspond to the same player.

\begin{figure}[t]
	\centering
	\includegraphics[scale=0.35]{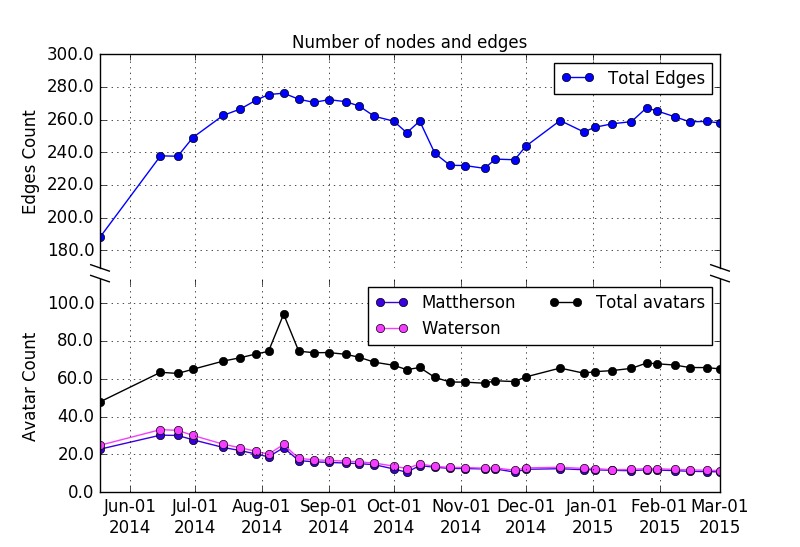}
	\caption{The population of avatars by origin.}
	\label{fig:MergePop}
\end{figure}

Before we continue, let us remark that all servers follow similar population trends to those shown in Figure \ref{fig:MergePop}. After spiking in September, the population declines through October to mid-November and then rises somewhat through December,  with another spike in the first month of the new year. 

\section{Network properties}
\begin{figure}%[H]
    \centering
    \includegraphics[scale=0.35]{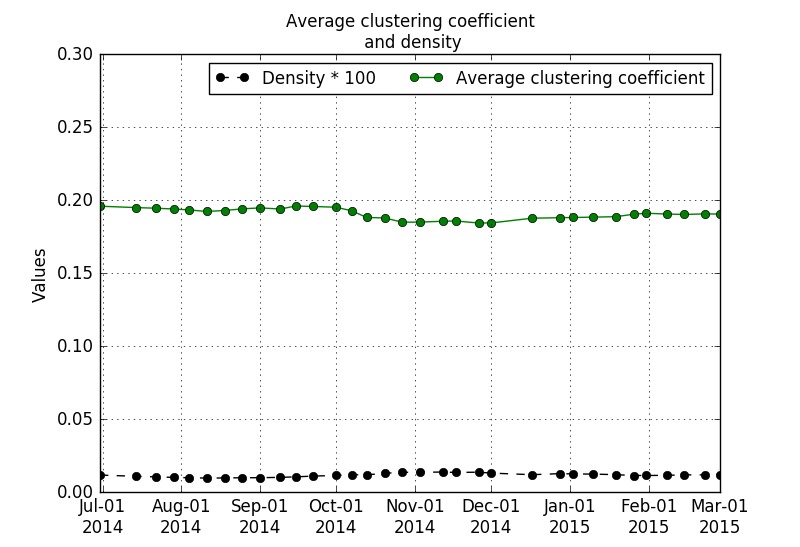}
	\caption{Average clustering coefficient  and density (multiplied by 100) as a function of time.}
	\label{fig:density_clust}
\end{figure}

The game networks exhibit power law degree distributions, high clustering coefficient, and large diameters. 
The average clustering coefficient and density of each snapshot are shown in Figure \ref{fig:density_clust}. Note that the density values are multiplied by 100 to allow comparison on the same scale.  Since the majority of avatars have a small degree and no mutual friends whatsoever, the average clustering coefficient of the snapshots is typically around 0.2. Clearly, the average clustering coefficient of the Planetside 2 dataset is much larger than the density, meaning that these networks exhibit high clustering among their nodes, as expected in a social network \cite{kadushin_understanding_2012}.

\begin{figure}%[H]
    \centering
    \includegraphics[scale=0.35]{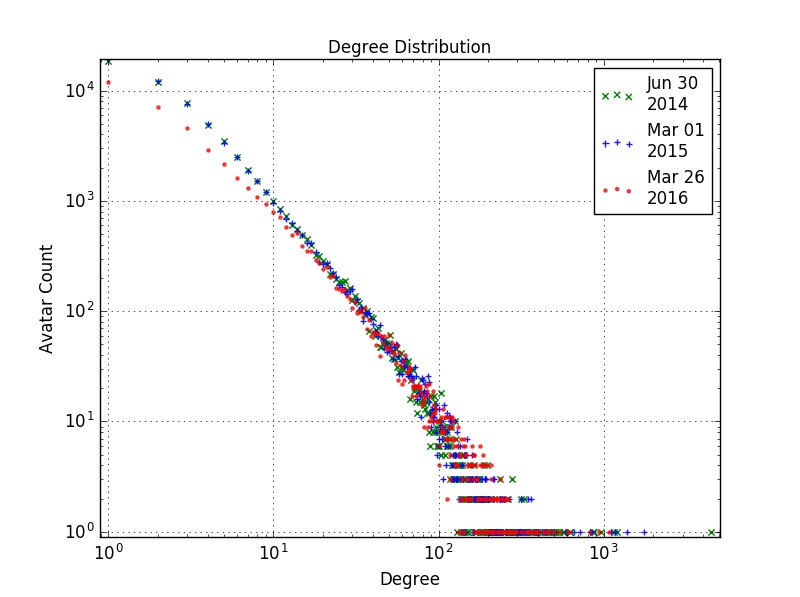}
    \caption{Typical degree distribution}
    \label{fig:EWdegDist}
\end{figure}

Figure~\ref{fig:EWdegDist} shows typical degree distributions of the network, using three snapshots as examples. Although 
the distributions appear to be linear in the log-log  plot, this is not a sufficient ``proof'' that the degree distribution 
has the form of a power law. In order to produce a more convincing evidence of the power law behavior, we used the method described in~\cite{powerlaw}. 

We first produced a best fit to degree distribution data using three models, the power law, the exponential function, and 
the power law with exponential cutoff.
The best fit for each of the three models was calculated by minimizing the Kolmogorov-Smirnov statistic,  $D = \max_{x \geq xmin} |S(x) - P(x)|$, where $S(x)$ is the CDF of the data, and $P(x)$ is the CDF of the model \cite{powerlaw}.  We then compared the quality of the resulting best fits using the log likelihoods. This revealed that the  exponential model is a poor fit for the data, meaning that the degree distribution is indeed best described by the power law.

As for the diameter, we found that the Planetside 2 snapshots have rather large diameters for social networks, typically between 13 and 15. This is a result of  a combination of factors.  The network consist of three faction with a  structure of long trees composed of peripheral avatars that  can have a depth of 4 or 5 in rare cases. Even within the same faction, the diameter is usually 13 and never less than 10. These large diameters are also found in both control servers, so they are not a result of the server merger.

\begin{figure}%[H]
    \centering
    \includegraphics[width=7cm]{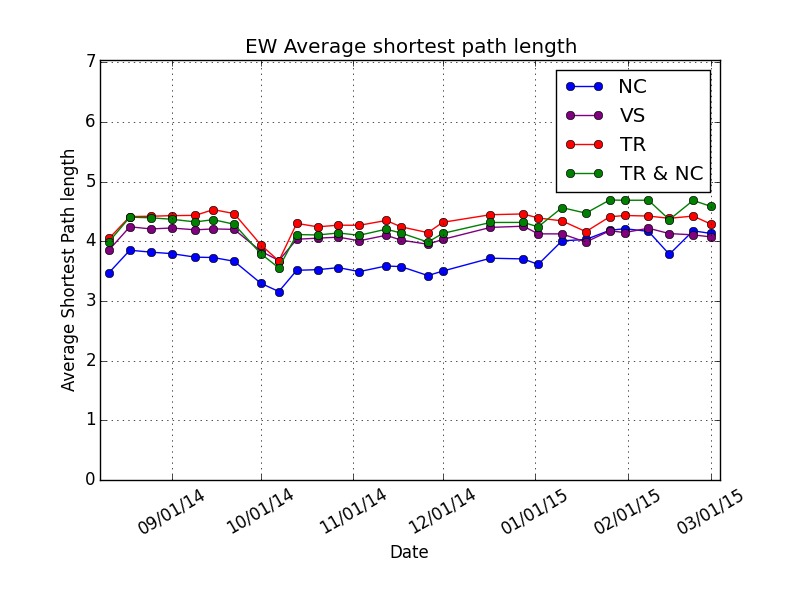}
    \caption{The average shortest path length of each of the three factions.}
    \label{fig:EW_diameter_APL}
\end{figure}

Figure \ref{fig:EW_diameter_APL} shows  the average path length (APL) of the giant component for each of the factions.\footnote{The highest APL always corresponds to the faction with the worst reputation on its server.}
An odd phenomenon can be observed in this figure. Until January, the combined TR \& NC faction had a lower APL than the TR alone. This can be
explained  as the influence of a single NC avatar. Once he stopped logging in in January, the APL of the TR \& NC became greater then that of the TR faction alone.  We verified this explanation by removing this avatar from  earlier snapshots.

\section{Server Merger}
The merger of the two US west coast servers, Mattherson and Waterson, was announced in the early spring of 2014, and took place late in July creating new the server named Emerald. The first snapshots was captured by our crawler on the 18th of May. 

Each node in the merged network is assigned an origin based on the server and faction it was originally created on. Nodes which have never been seen on either of the original servers are called \emph{newcomers}. It is important to note that the set of newcomers includes  both the newly created avatars and avatars who have not been active since early February. Ideally, returning avatars should be placed on the correct server, but this was not possible. However,  with a few exceptions, these returning avatars had mostly stopped playing soon after they were created, therefore, for all practical purposes, they are newcomers to the current social context.

\begin{figure*}[htbp]
    \subfloat[Waterson]{\includegraphics[width=\textwidth/2]{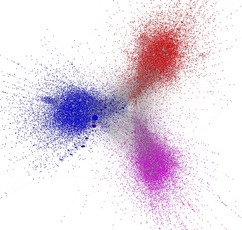}}
    \subfloat[Mattherson]{\includegraphics[width=\textwidth/2]{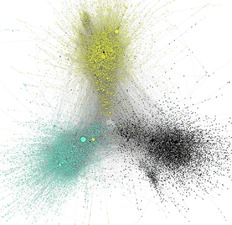}}
    \caption{Visualisation of Waterson and Mattherson server networks on May 18 (before the merger).}
    \label{fig:May18}
\end{figure*}

In what follows, we will show a number of visualizations illustrating the merger and the subsequent evolution of the resulting network.
These visualizations were created using Gephi's ~\cite{ICWSM09154} Force Atlas 2 layout, using default parameters. Each avatar is assigned a colour according to Table \ref{tab:colourKey}. The size of the avatar scales linearly with its degree,  except the highest degree avatar, whose size is limited to twice that of the second largest.
\begin{table}%[H]
	\centering
    \begin{tabular}{r!{\color{white}\vrule} c!{\color{white}\vrule}  c!{\color{white}\vrule} c}
        Origin & \multicolumn{3}{c}{Faction} \\
         & NC & TR & VS\\
        \arrayrulecolor{white}\hline
        Waterson & \cellcolor{WNC} & \cellcolor{WTR} & \cellcolor{WVS}\\[1.4em]
        \arrayrulecolor{white}\hline
        Mattherson & \cellcolor{MNC} & \cellcolor{MTR} & \cellcolor{MVS}\\[1.4em]
        \arrayrulecolor{white}\hline
        Newcomer & \cellcolor{NNC} & \cellcolor{NTR} & \cellcolor{NVS}\\[1.4em]
    \end{tabular}
    \caption{Colour key for avatars.}
    \label{tab:colourKey}
\end{table}
Pre-merger snapshots made on the May 18th are provided in Figure \ref{fig:May18}. Each of three lobes corresponds to a faction. When considered separately, each individual lobe is a social network. The distinctive three-lobed structure is common to all the unmerged servers that we looked at (Connery and Miller). After some time the newly formed Emerald will resemble them as well. 
%%%%%%%%%%%
\begin{figure*}%[H]
\centering
%%%%%%%%%%%
\begin{minipage}[b]{\textwidth-5cm}
\subfloat{\includegraphics[width=\textwidth]{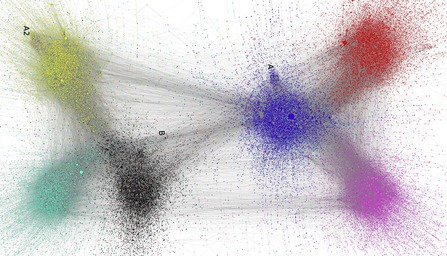}}
\end{minipage}
\qquad
\begin{minipage}[b]{4cm}
\centering
\vfill
\subfloat{\includegraphics[width=4cm]{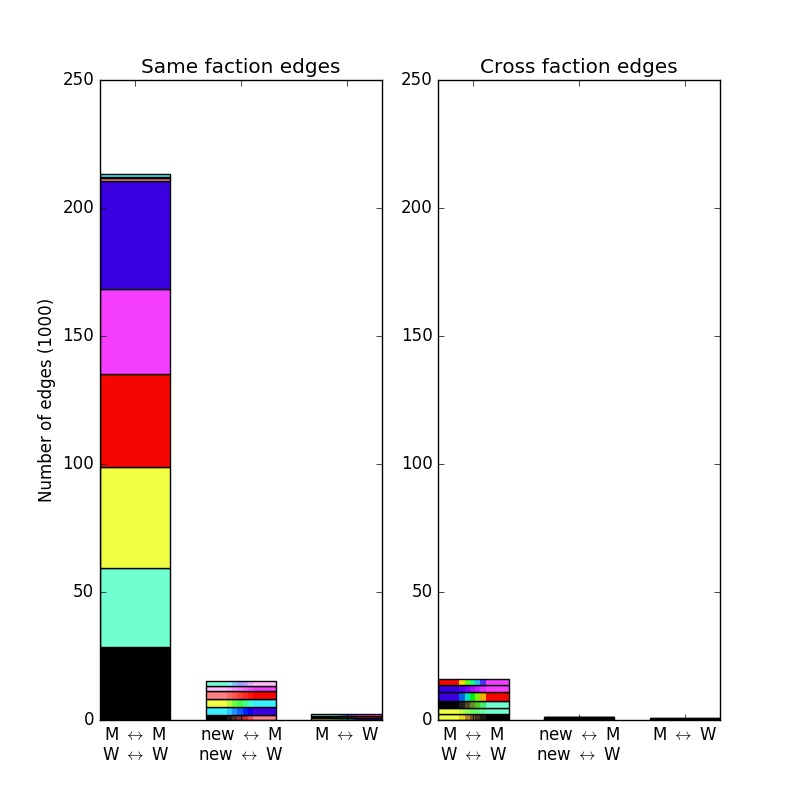}}
\vfill
\subfloat{\includegraphics[width=4cm]{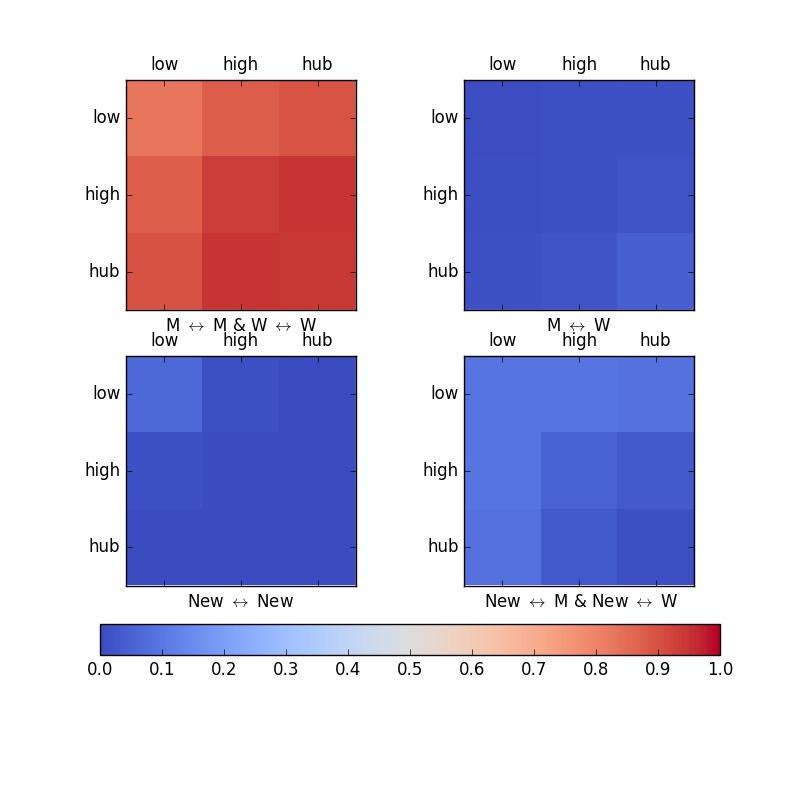}}
\end{minipage}
\caption{Visualisation of Emerald server network after the merger (on June 30th).}
\label{fig:June30}
\end{figure*}

The first snapshot of the new Emerald server was captured on the 30th of June, and it is visualized in Figure \ref{fig:June30}. 
It is accompanied by two figures that illustrate how the edges are connected. The bars graphs break down edges of the network
according to the origins of the nodes which the edge connects, where $M$ and $W$ labels denote, respectively,  nodes 
originating from Mattherson and Waterson servers. Histogram use the same colours as the visualizations.
 Two bar graphs are provided, one for faction edges and another for cross-faction edges. Cross-faction edges which are further broken down by the server origin. The second figure (right bottom of Figure \ref{fig:June30}) shows  ``heat maps'' illustrating how these connections connect avatars of different degrees. Low degree avatars have a degree less than or equal to 8, which is the average degree.  High degree avatars have degrees between 8 and 96, and those with still higher degrees are called \emph{hubs}. Around half of all edges connect to a low degree node, while high and hub nodes split the remaining edges. Note that the entries of the matrices in ``heat maps'' are normalized by the overall number of edges in each element in the entire network.

Analysis of the first post-merger snapshot shows that already 11\% of the population consists of newcomers.  As we can see in  edge breakdown histograms, few cross server connections have had time to form by this point, so the original networks are clearly distinct. However, significant number of edges have already formed between low degree newcomers and the originals.

Labels $A$ and $A2$ in in Figure \ref{fig:June30} mark unusually insular outfits with strong internal connections. Since they have unusually few connections to their faction's core hubs, they haven't been pulled into the core of the network like most other outfits.  
Label $B$ marks a YouTube celebrity in the game community, which is immediately pulled into a position between the two servers due to cross server edges that predate the merger. This avatar functions as a broker \cite{kadushin_understanding_2012,shen_virtual_2012}.

\begin{figure}%[H]
	\centering

	\includegraphics[width=\textwidth/2]{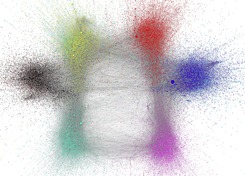}
	\includegraphics[width=4cm]{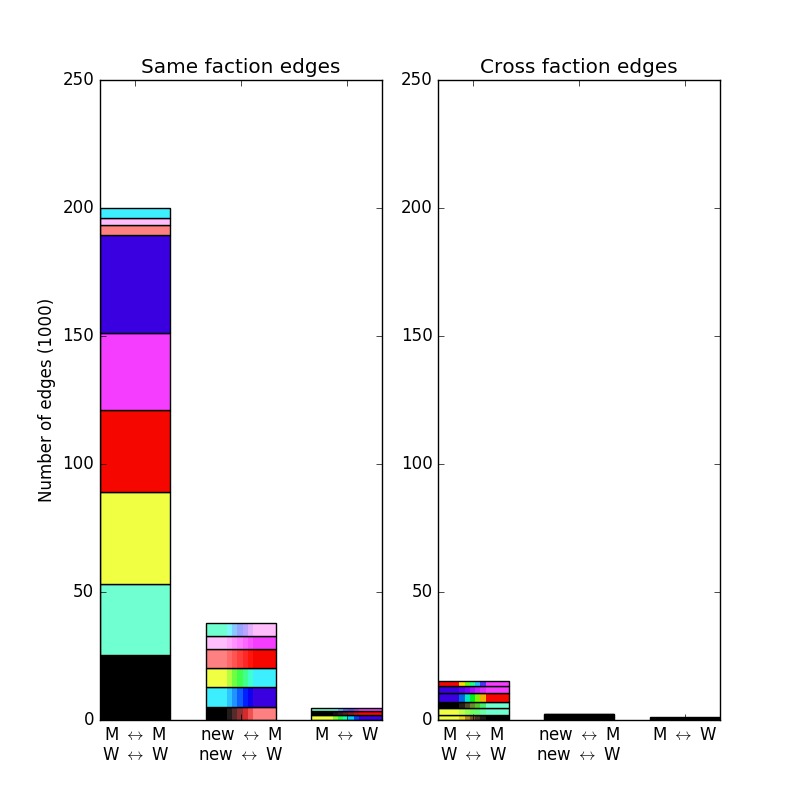}\includegraphics[width=4cm]{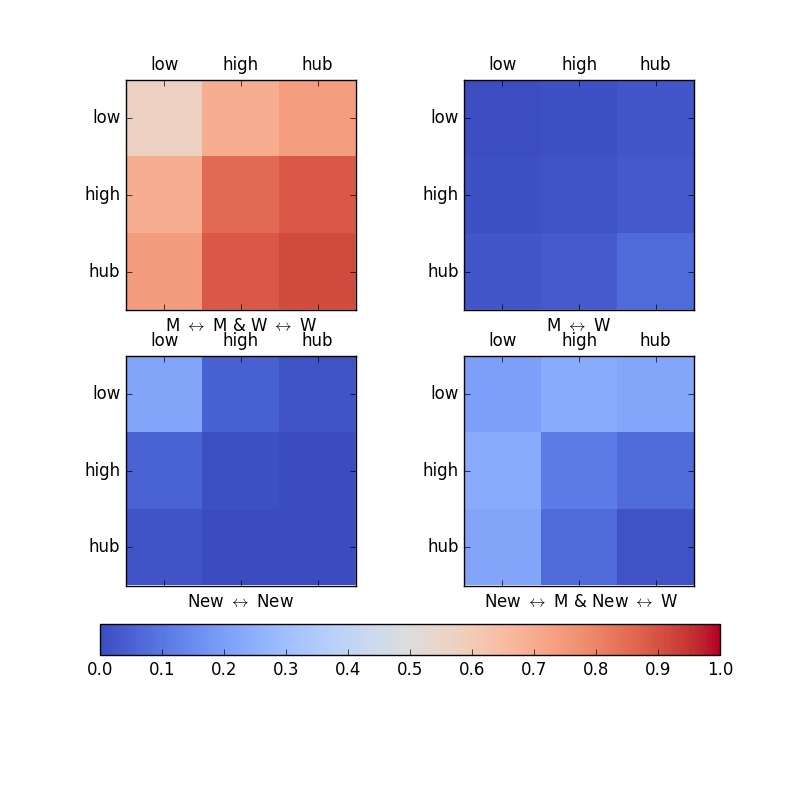}
	\caption{Emerald server network on July 14th.}
	\label{fig:July14}
\end{figure}

The second snapshot, form June 14th, is shown in Figure \ref{fig:July14}. By this point, 28.8\% of the population are newcomers which have begun to replace the original periphery. As we can see, the cross faction edges among the original Mattherson and Waterson factions are preventing the combined factions of Emerald from coming together.

%The original cross faction edges are the edges that connect the three factions from both original servers and they are currently holding the new Emerald factions apart. 

%\begin{figure*}%[H]
%\centering
%\begin{minipage}[b]{\textwidth-5cm}
%\subfloat{\includegraphics[width=\textwidth]{MergerPics/smallEmerald July 14 v2.jpg}}
%\end{minipage}
%\qquad
%\begin{minipage}[b]{4cm}
%\centering
%\vfill
%\subfloat{\includegraphics[width=4cm]{ServerMerger/July14.jpg}}
%\vfill
%\subfloat{\includegraphics[width=4cm]{DegDiffPics/July14.jpg}}
%\end{minipage}
%\caption{Emerald server network on July 14th.}
%\label{fig:July14}
%\end{figure*}

\begin{figure}%[H]
    \centering
    \includegraphics[width=\textwidth/2]{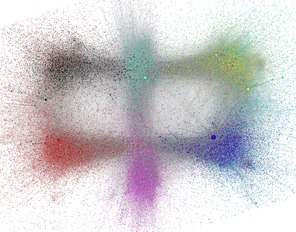}
    \includegraphics[width=4cm]{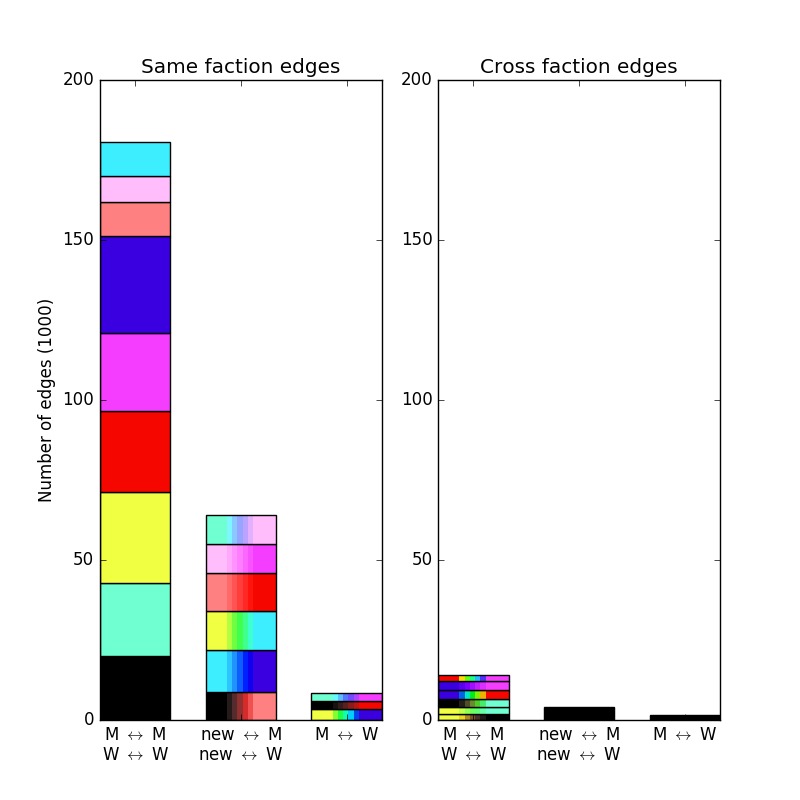}\includegraphics[width=4cm]{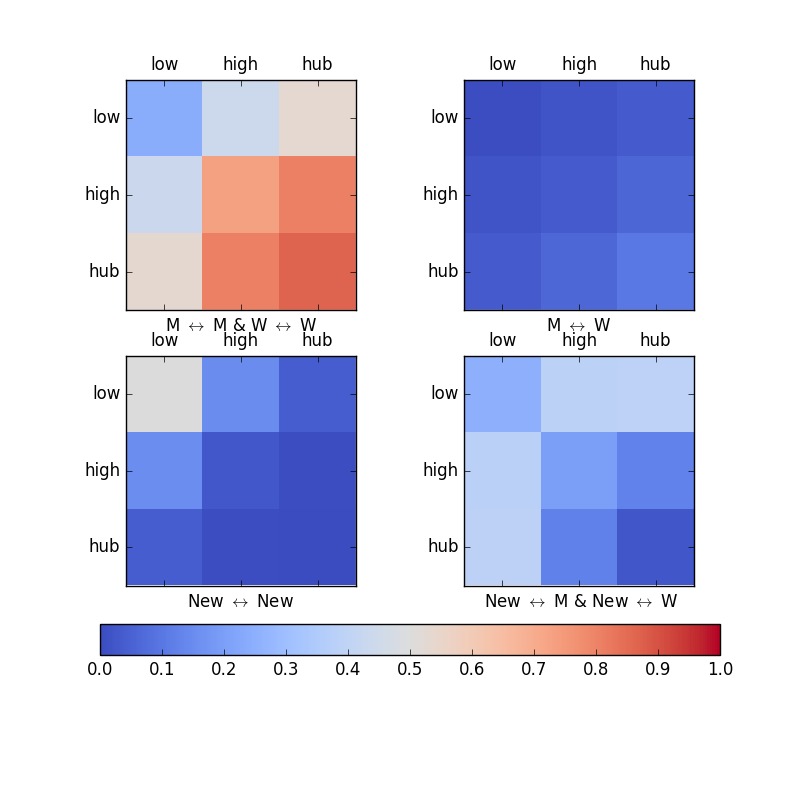}
    \caption{Emerald server network on Aug 18th.}
    \label{fig:Aug18}
\end{figure}
By the 18th of August, the combined indirect and direct $M \leftrightarrow W$ edges have overcome the combined number of the original
 cross-faction edges. There are only half as many direct edges as original cross-faction edges.
A number of  indirect edges running through newcomers (as shown in Figure \ref{fig:directvsindirect}) has overcome the original servers' structure and the three factions have begun to merge.
\begin{figure}%[H]
\centering
    \includegraphics[width=\textwidth/2]{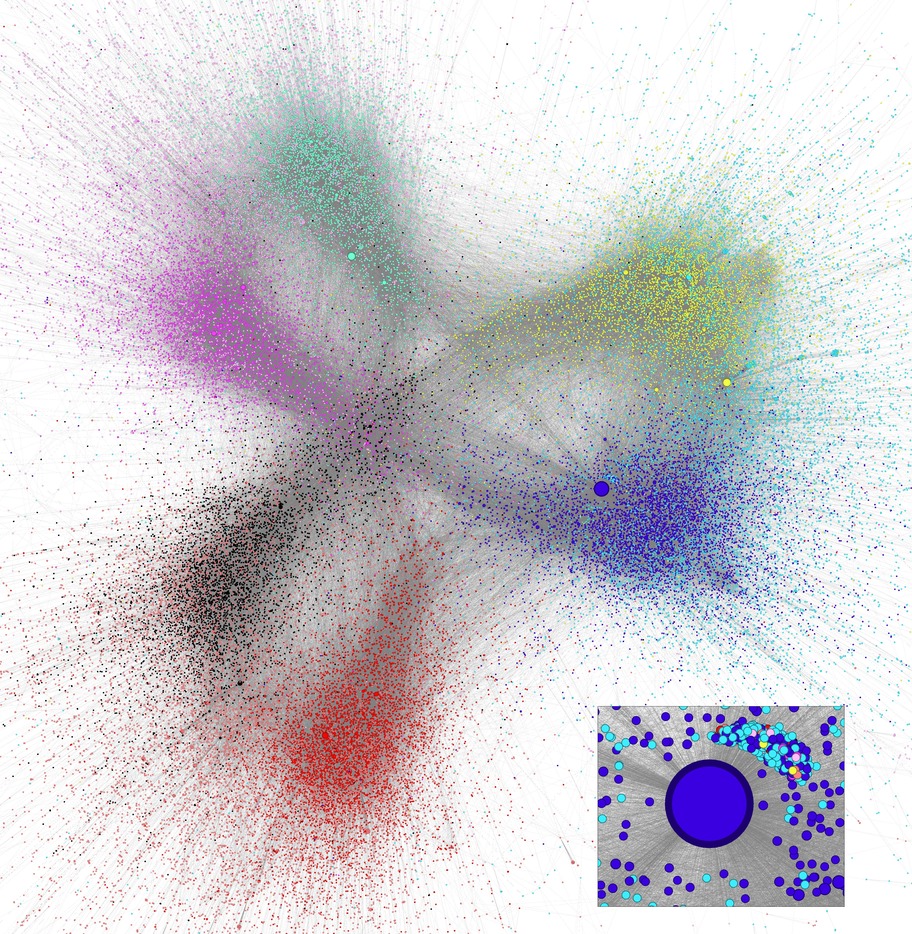}
    \includegraphics[width=4cm]{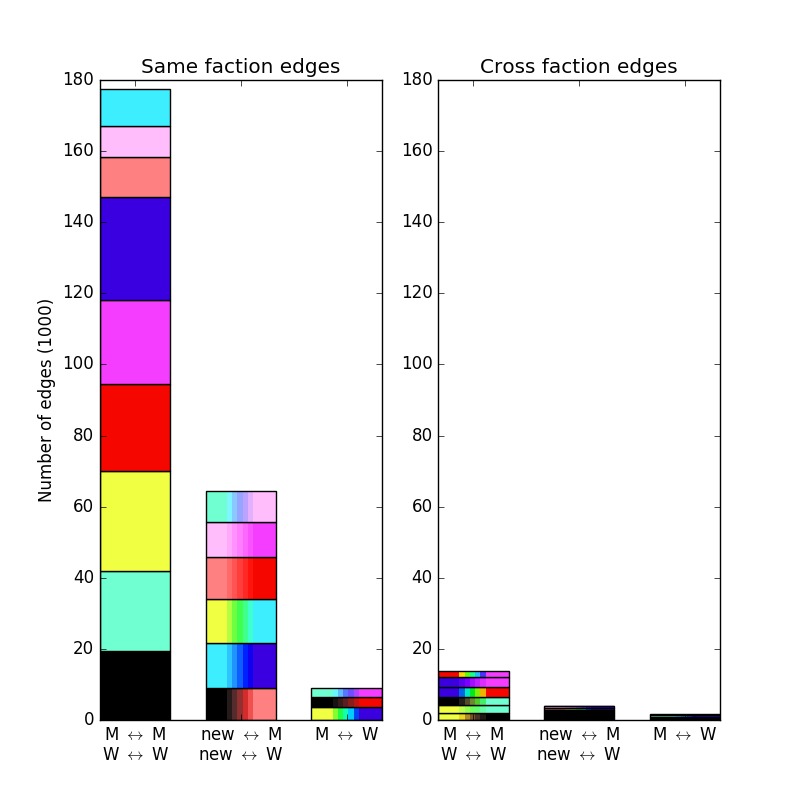}\includegraphics[width=4cm]{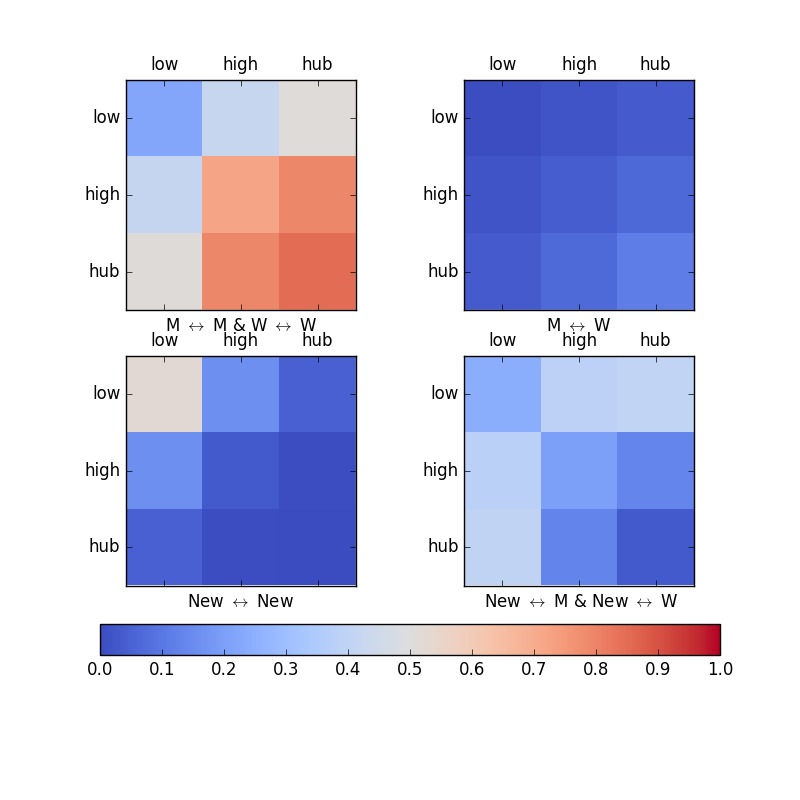}
    \caption{Emerald server network on  August 24th.}
    \label{fig:Aug24}
\end{figure}
It only takes one more week to return to the familiar three lobed structure, and we can see it in Figure \ref{fig:Aug24} recorded on the 24th of August. Clearly, the newcomers are either between the original cores or replacing the original peripheral avatars. We also show expanded  area around Klypto, the highest degree avatar, which has over 5100 friends in this snapshot. In fact, Klypto has so many edges to Waterson TR avatars that it visibly deforms the shape of the combined TR lobe. 

% Removeing september 15 seems harmless. 

%\begin{figure}%[H]
%    \centering
%    \includegraphics[width=\textwidth/2]{MergerPics/smallEmerald Sept 15.jpg}
%    \includegraphics[width=4cm]{ServerMerger/Sept15.jpg}%\includegraphics[width=4cm]{DegDiffPics/Sept15.jpg}
%    \caption{Emerald server network on September 15.}
%    \label{fig:Sept15}
%\end{figure}
%Over the next month this trend continued. In Figure \ref{fig:Sept15} from the 15th of September, we see that the original peripheral avatars have been mostly replaced with newcomers. 

This trend continued through the month of September, with the original servers becoming the minority as the original peripheral avatars are replaced with newcomers. 

\begin{figure}[t]
    \centering
    \includegraphics[width=\textwidth/2]{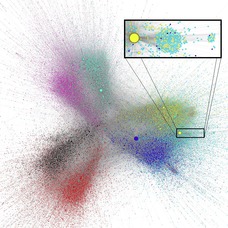}
    \includegraphics[width=4cm]{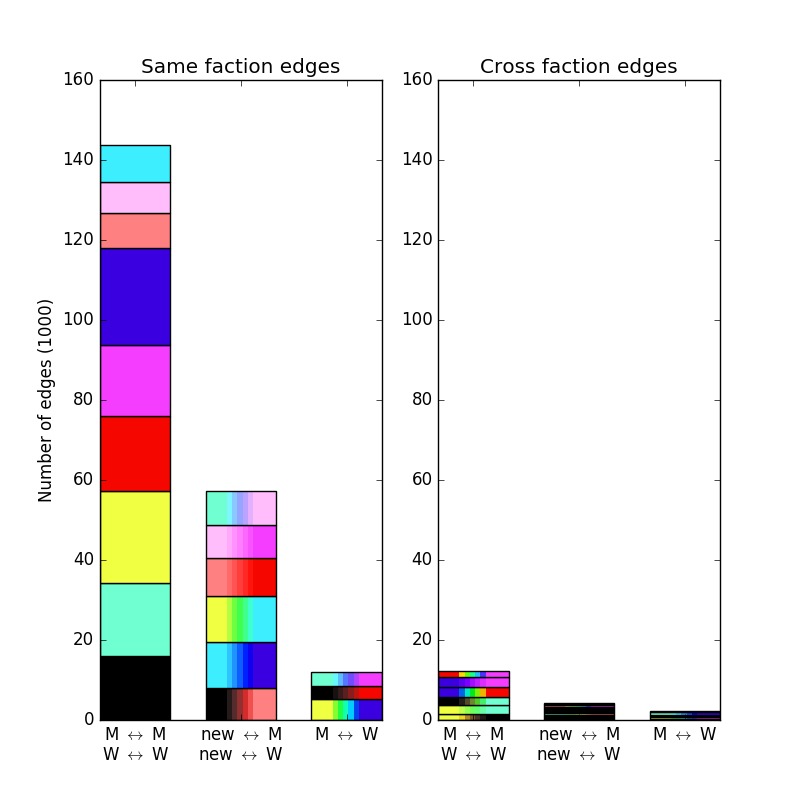}\includegraphics[width=4cm]{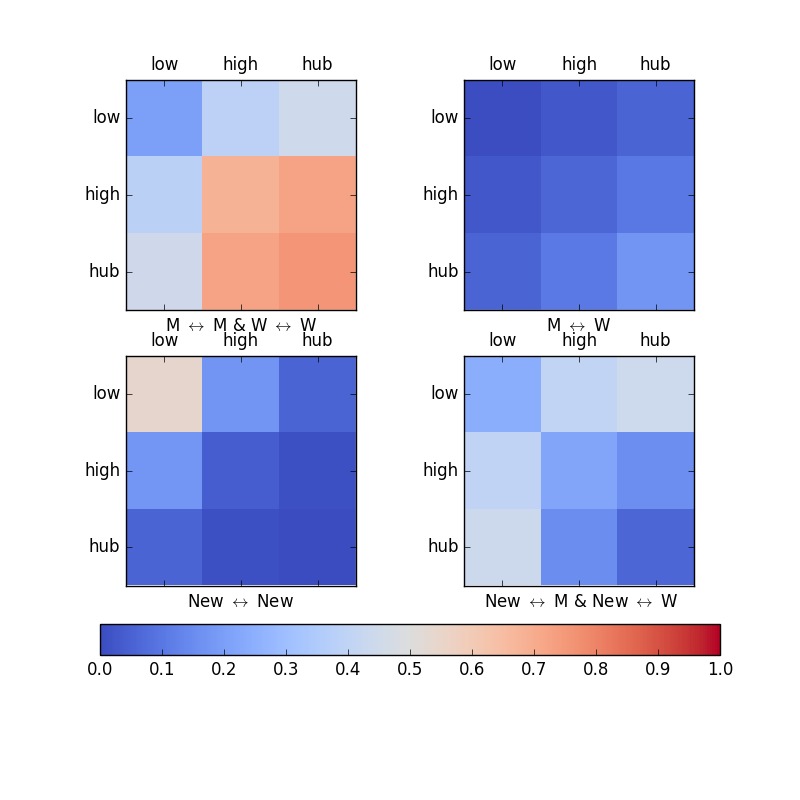}
    \caption{Emerald server network on  October 27th.}
    \label{fig:Oct27}
\end{figure}

In Figure \ref{fig:Oct27} we have zoomed in on the avatars who are connected to the second highest degree avatar. The avatars in the center have no neighbors other than the hub, those in the right cluster are all connected to exactly one other friend in the same cluster. These long trails of low degree avatars are common throughout the snapshots.

\begin{figure}[t]
    \centering
    \includegraphics[width=\textwidth/2]{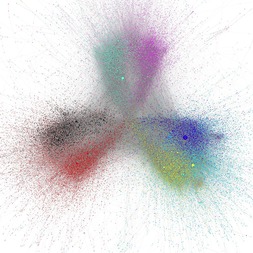}
    \includegraphics[width=4cm]{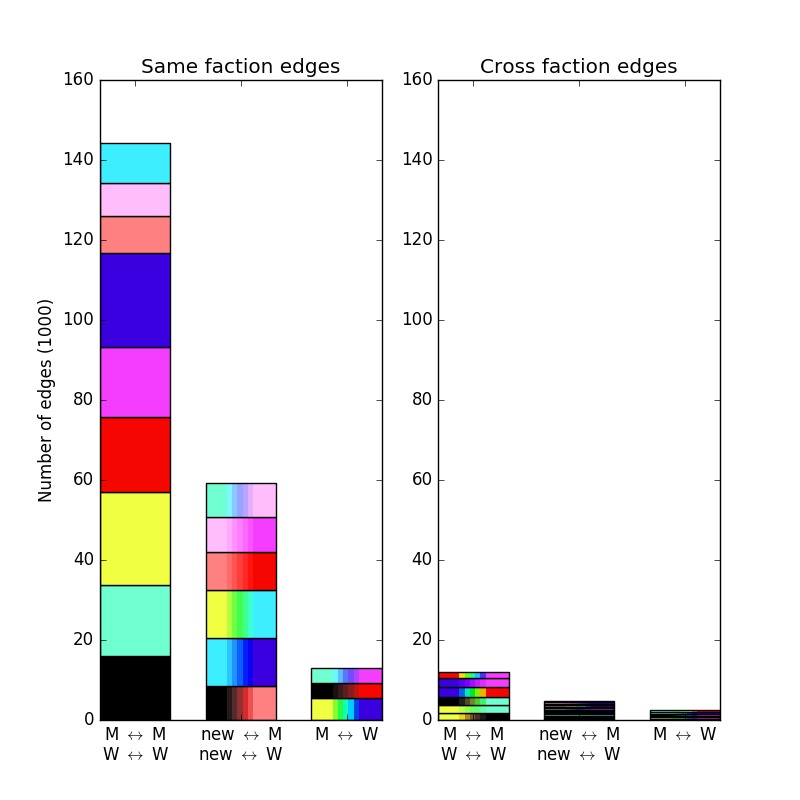}\includegraphics[width=4cm]{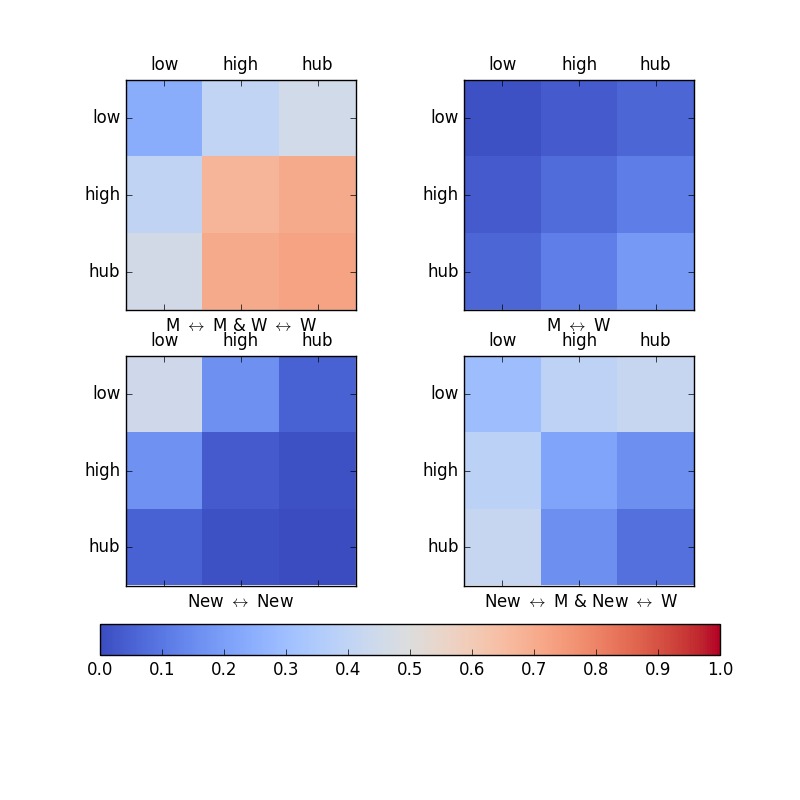}
    \caption{Emerald server network on November 17th.}
    \label{fig:Nov17}
\end{figure}

By November 17th (see Figure\ref{fig:Nov17}), the merger is nearly complete, with most of the peripheral avatars replaced by newcomers who now make up two-thirds of the population. The rate at which new direct edges are formed between the two original groups has been steady for over a month (Figure \ref{fig:directvsindirect}), while the average degree has also stabilized (Figure \ref{fig:average_degree_by_original_server}). 

Visually, it is now difficult to distinguish the original Mattherson or Waterson servers without the aid of colours. However, 
when we look at the assortativity in Figure \ref{fig:Merger_assortivity} (to be further explained in in sec. V), we can see that the divisions are clearly still there.

\begin{figure}[t]
	\centering
    \includegraphics[width=\textwidth/2]{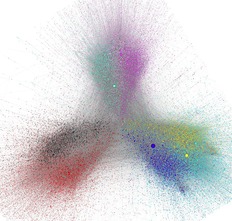}
    \includegraphics[width=4cm]{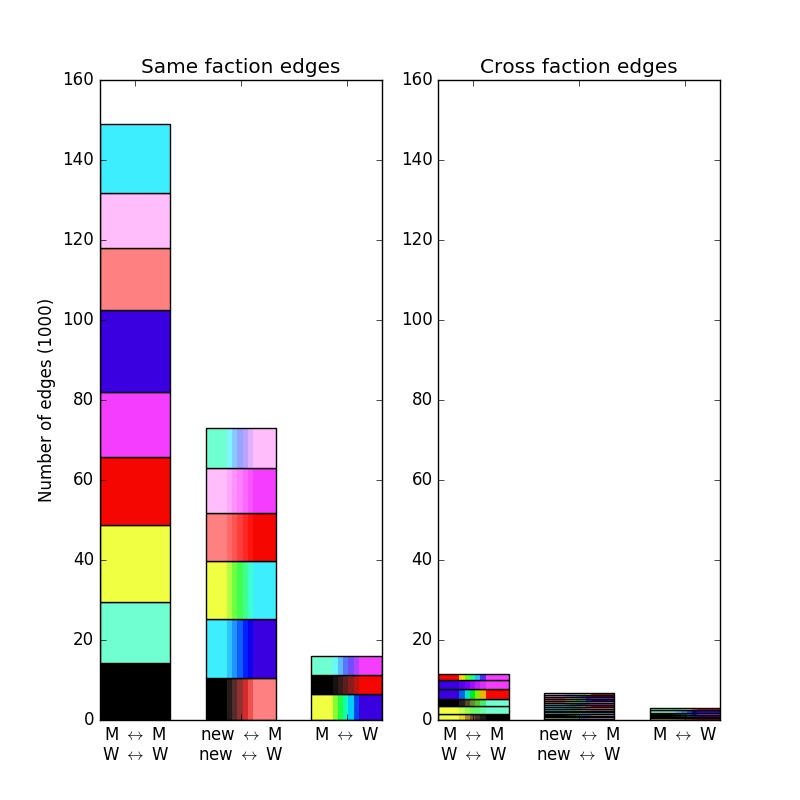}\includegraphics[width=4cm]{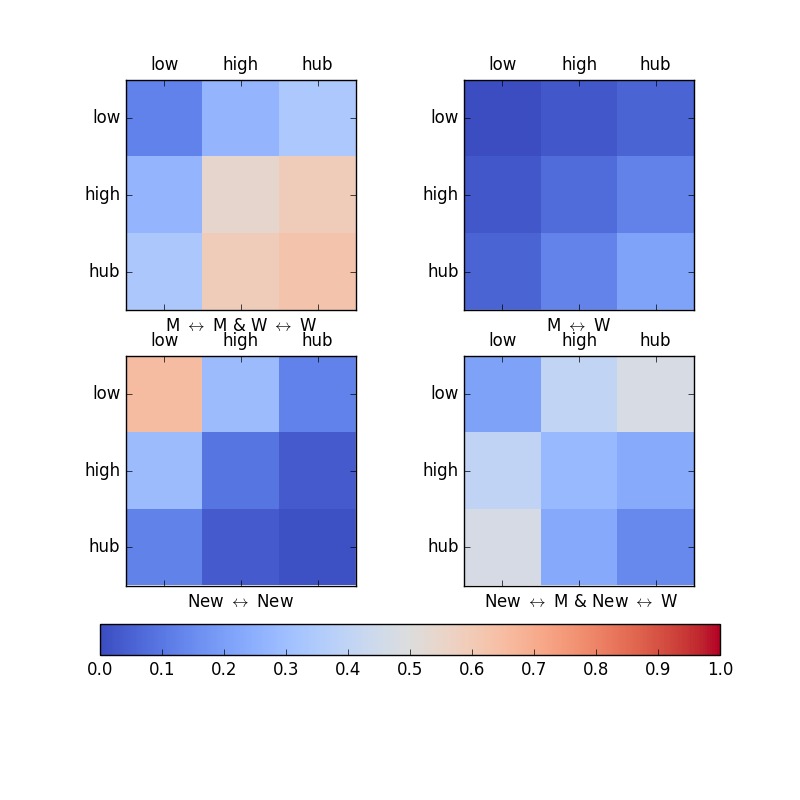}
    \caption{Emerald server network on February 23rd.}
	\label{fig:Feb23}
\end{figure}

 Figure \ref{fig:Feb23} shows the last snapshot from 2015. The number of edges between the Mattherson and Waterson avatars still do not outnumber those connecting them to themselves. Even more interestingly, the majority of edges are incident to the original avatars.  
The number of avatars remaining from the original servers dwindled down to about 6\% of the population.

\begin{figure}[t]
    \centering
    \includegraphics[width=\textwidth/2]{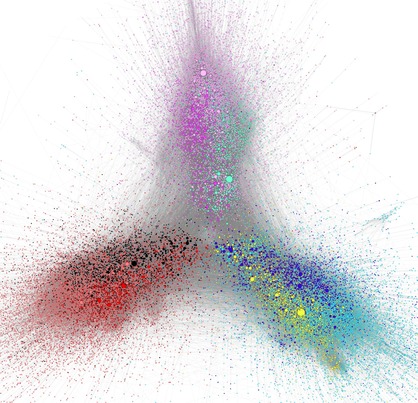}
    \includegraphics[width=4cm]{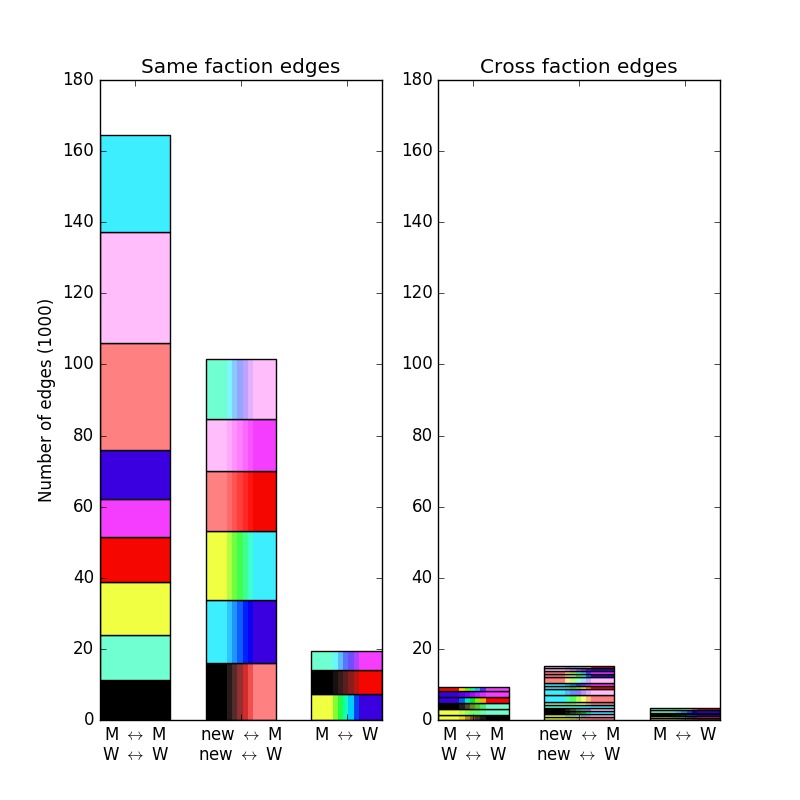}\includegraphics[width=4cm]{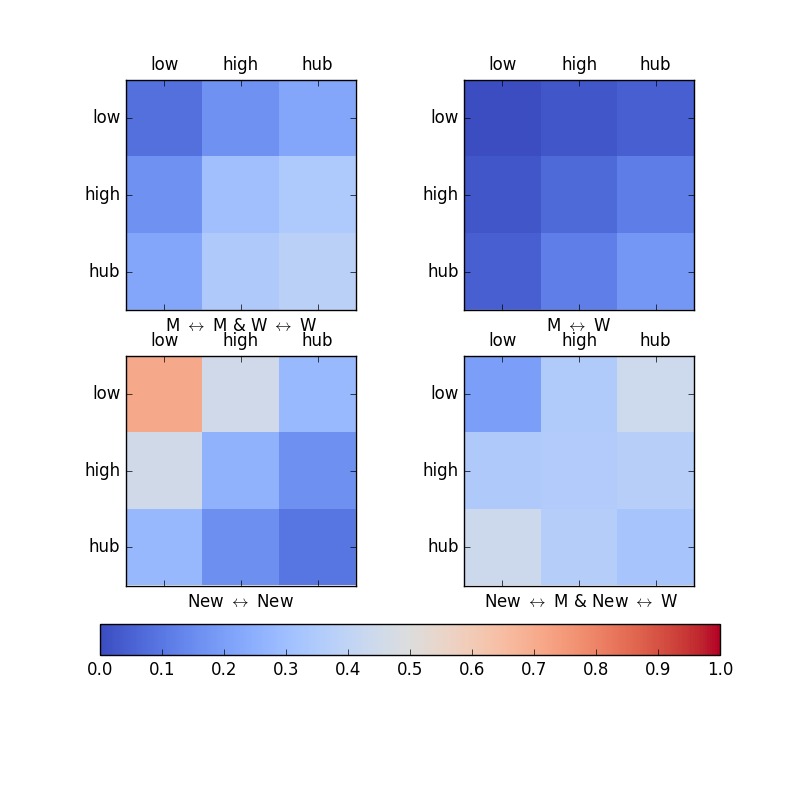}
\caption{Emerald server network on March 26, 2016}
    \label{fig:Mar26_2016}
\end{figure}

The final snapshot (Figure \ref{fig:Mar26_2016})  shows the server just over a year and a half later, on the 26th of March 2016. The server structure has changed little. The two separate cores in each faction have decreased to around 4.6\% of the population of each, while the number of avatars has decreased to levels seen in the pre-merger servers.

\section{additional measurements}
In this section we examine the merger with some additional measurements.
Assortativity, as described in ~\cite{newman_mixing_2003}, is the tendency of nodes to be more often connected to nodes similar or dissimilar to themselves in some way. The assortativity coefficient reflects how strongly the members of groups tend to stick to their own.

First let $e_{i,j}$ be the fraction of edges that connect type $i$ and type $j$ vertices in a graph $G$. Let $a_{i}$ be the fraction of edges connecting to type $i$ verticals. The assortativity coefficient $r$ is defined as
$$r = \frac{\sum_{i} e_{i,i} - \sum_{i} a_{i}^2}{1 - \sum_{i} a_{i}^2}.$$
A high value of  $r$ indicates that the avatars prefer to stick to connecting to other members of their group in that partition.

\begin{figure}
    \centering
    \includegraphics[scale=0.35]{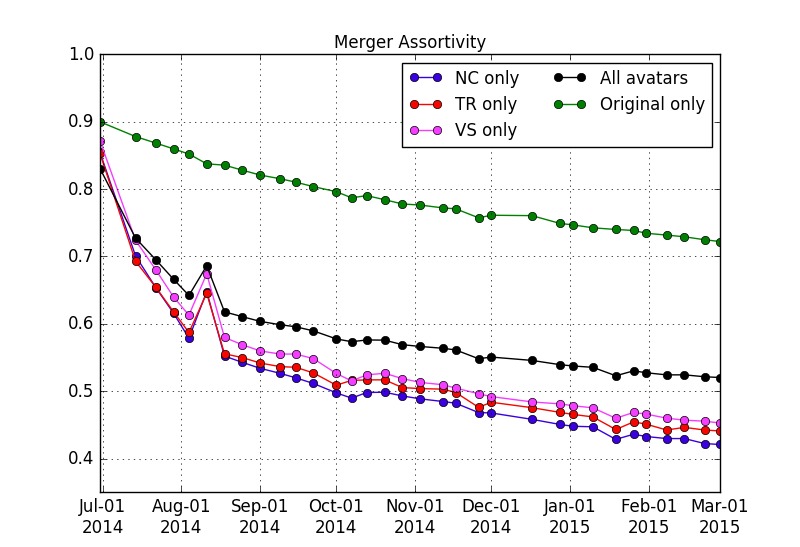}
	\caption{The assortativity by origin.}
	\label{fig:Merger_assortivity}
%\end{figure}    

%\begin{figure}
	\centering
	\includegraphics[scale=0.35]{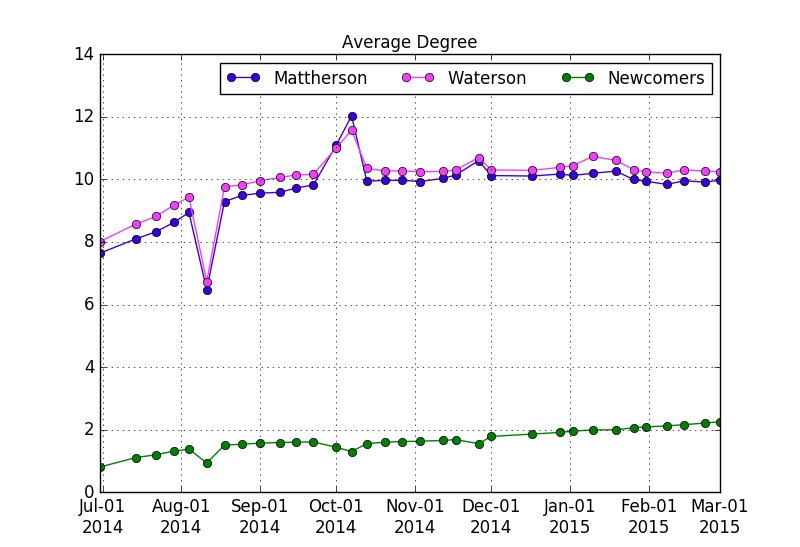}
	\caption{Average degree by origin.}
	\label{fig:average_degree_by_original_server}
%\end{figure}

%\begin{figure}
    \centering
    \includegraphics[scale=0.35]{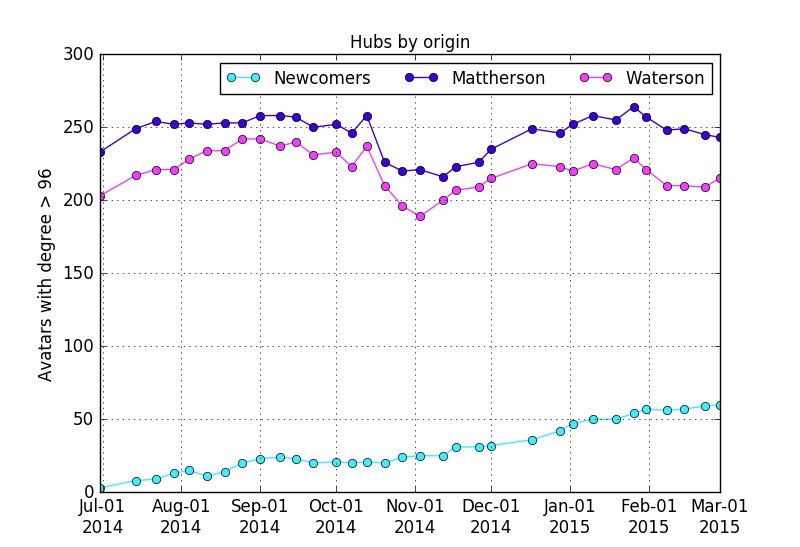}
    \caption{Breakdown of high degree avatars by origin.}
    \label{fig:hub_breakdown}
%\end{figure}

%\begin{figure}
	\centering
	\includegraphics[scale=0.35]{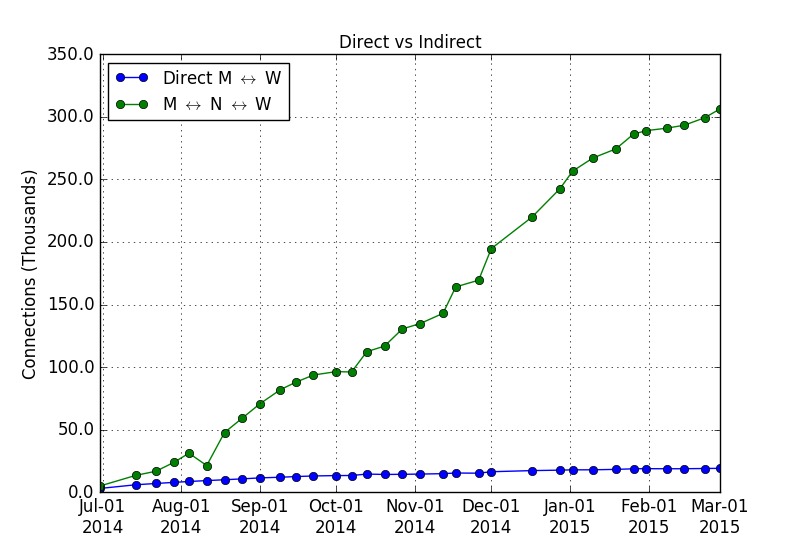}
	\caption{The number of direct edges between Mattherson and Waterson nodes versus the paths through newcomers.}
	\label{fig:directvsindirect}
\end{figure}

The  assortativity coefficient of every snapshot when avatars are grouped by origin are shown in Figure \ref{fig:Merger_assortivity}. ``All avatars'' line represents all nine origins, while the ``original only'' line  represents  the assortativity with newcomers excluded. The  NC, TR, and VS lines show the assortativity between origins within each respective faction.  

The assortativity between the originals only is the highest since the two original servers tend to mix the least. With newcomers included, the assortativity is much lower since all groups are mingling together much more as they mutually bond to newcomers. Each individual faction's assortativity is likewise lower since of course they prefer to form bonds to teammates over enemies. The NC as the faction with the highest degree avatar shows the lowest assortativity of all. Nevertheless,  these are very high assortativities, and they all change in step with each other.

As shown in Figure \ref{fig:average_degree_by_original_server}, the average degree of the avatars from Mattherson and Watterson grows as peripheral low degree avatars are replaced with newcomers. The average degree of the newcomers begins to rise as new hubs begin to take 
their places between factions' cores, as seen in Figure \ref{fig:hub_breakdown}.

The number of hub avatars by origin as a function of time is given in Figure \ref{fig:hub_breakdown}. Recall that we define hubs to be avatars with 96 or more active friends\footnote{96 is the number of avatars in two 48-player platoons, and close to the minimum degree of an avatar in the heavy tail of the degree distribution~\cite{powerlaw}.}. Typically, there are approximately 500-600 such avatars in any given snapshot. The number of hubs from the original servers is pretty consistent but as time passes the original hubs are joined by the growing newcomer hubs. Unlike the peripheral nodes, newcomer hubs join  existing hubs rather then replacing them. This is because, in general, hubs do not tend to leave, and those that do tend to return soon.

The Figure \ref{fig:directvsindirect} compares the number of direct links between the Mattherson and Waterson avatars to the number of indirect connections through newcomers. The indirect links are clearly the dominant force binding the two networks together. The rate at which both kinds form is very consistent over time and the bridging role of the newcomers is very clear.

The slow formation of direct edges is especially interesting since there is no way to tell  which server any given avatar is actually from. The only way to find out where someone is from in the game would be to ask, or to make a guess based on the outfit. However, the majority of avatars are not in an outfit.

\section{Conclusions and future work}
The presented evidence shows that it takes a surprising amount of time for the populations of two servers to 
actually start to mingle after the merger of servers. Seven months, after all,  is quite a long time in the life span of a video game community. With a few notable exceptions, the lifespans of most games and their associated communities are not longer than five years.
With nearly 70\% of the population of the server being replaced by newcomers after seven months, one can say that the decay of the original structures was more important  in this process that the growth of interconnections between original avatars.

The lack of direct edges between avatars from pre-merger servers is especially interesting given that there is no way to tell  which server any given avatar is from. However, one should keep in mind that 60\% of all avatars are not in an outfit at all, and that both of the original servers were in the same time zone and their players belonged to similar cultures.  This makes the phenomenon even more remarkable.
One could even say that the actual social cores of the original two networks have not actually ``merged'', but rather  their core avatars have mutually bonded with newcomers, who replaced the original peripheral avatars.

Our observations are mostly phenomenological, but since there exists a lot of models for dynamic networks \cite{hanneke_discrete_2010,davidsen_emergence_2002,welles_dynamic_2014}, it would be interesting to see how these
models behave under the conditions of a merger. Work in this direction is planned in the near future.

\textbf{Acknowledgements}: H.F. and B.F. acknowledge financial support from the Natural Sciences and
Engineering Research Council of Canada (NSERC) in the form of Discovery Grants.
%\section{Figures should not be hiding in the citations...}
%\clearpage
\bibliographystyle{IEEEtran}
\bibliography{asonampaper}
\end{document}